\documentclass{article}

\usepackage{latexsym}

\usepackage{amssymb}

\newcommand{\fzero}{f^{\mathrm{in}}}

\def\prfe{\hspace*{\fill} $\Box$

\smallskip \noindent}
\newtheorem{Theorem}{Theorem}

\newtheorem{Lemma}{Lemma}                                                           

\DeclareFontFamily{OT1}{rsfs}{}                                                 
\DeclareFontShape{OT1}{rsfs}{m}{n}{ <-7> rsfs5 <7-10> rsfs7                     
<10->rsfs10}{}                                                                  
\DeclareMathAlphabet{\mycal}{OT1}{rsfs}{m}{n}                                   
\newcommand{\scri}{{\mycal I}}

\title{Outgoing radiation from \\an isolated collisionless plasma}
\author{Simone Calogero\\[0.5cm]
Department of Mathematics, Chalmers University\\
S-412 96 G\"oteborg, Sweden\\
E-mail: mg026@math.chalmers.se}
\date { }
                                                                              
\begin{document} 
\maketitle

\begin{abstract}
The asymptotic properties at future null infinity of the solutions of the relativistic
Vlasov-Maxwell system whose global existence for small data has been established by the author in a previous
work are investigated. These solutions describe a collisionless plasma isolated from incoming radiation. 
It is shown that a non-negative quantity associated to the plasma decreases as a consequence of the dissipation of energy in form of 
outgoing radiation.
This quantity represents the analogue of the Bondi mass in general relativity.  
\end{abstract}

\section{Introduction and main results}
\setcounter{equation}{0}
The dynamics of a collisionless plasma in interaction with the mean electromagnetic field generated by the charges is described by the 
relativistic Vlasov-Maxwell system. In this model the unknowns are the electromagnetic field $(E,B)$ and a set of $N$ non-negative  
functions $f_\alpha$ which give the distributions 
in phase space of $N$ different species of particles. The system consists of the Vlasov equation
\[
\partial_tf_\alpha+\widehat{p}_\alpha\cdot\partial_xf_\alpha+q_\alpha(E+\widehat{p}_\alpha\wedge
B)\cdot\partial_{p}f_\alpha=0,\quad\forall\,\alpha=1,...,N,
\]
coupled to the Maxwell equations with charge density $\rho$ and current density $j$ given by
\begin{equation}\label{source}
\rho(t,x)=\int_{\mathbb{R}^3} \big( \sum_{\alpha=1}^{N} q_{\alpha}f_{\alpha}\big)\,dp,\quad
j(t,x)=\int_{\mathbb{R}^{3}} \big( \sum_{\alpha=1}^{N} q_{\alpha}f_{\alpha}\widehat{p}_{\alpha}\big)\,dp.
\end{equation}
In the previous equations, $t\in\mathbb{R}$ is the time, $x\in\mathbb{R}^3$, $p\in\mathbb{R}^3$ are the position and the momentum of the 
particles, $q_\alpha$ the charge of a particle of species $\alpha$, 
\[
\widehat{p}_{\alpha}=\frac{p}{\sqrt{m_{\alpha}^{2}+p^{2}}},\quad p^2\equiv |p|^2
\]
denotes the relativistic velocity and $m_\alpha$ is the mass of a particle of species $\alpha$. Units are chosen so that the speed of light 
is equal to unity. 

The relativistic Vlasov-Maxwell system has several applications in plasma physics and in astrophysics, where it is used for instance to model
the dynamics of the solar wind. Many mathematical problems remain unsolved. For example, existence of global classical solutions 
is known only under certain restrictions on the size of the initial data (see \cite{C, GS2, GSh1, R}); global existence for large data has been proved for a modified version of the system in which the particle
density is forced to have compact support in the momentum (see \cite{GS1}).

An important feature of the dynamics, which is due to its relativistic character, is the presence of radiation fields. 
The radiation is defined as the part of the electromagnetic
field which carries energy to null infinity. It is distinguished in \textit{outgoing} radiation, which propagates energy to the future null
infinity $\scri^+$, and \textit{incoming} radiation, which propagates energy to the past null infinity $\scri^-$. The latter can be
interpreted as a flux of energy flowing in onto the system from $\scri^-$. For an isolated system the incoming radiation should be ruled out
by appropriate boundary conditions, which will be now briefly discussed.

The amount of energy $\mathcal{E}_{\mathrm{in}}(v_1,v_2)$ carried to $\scri^-$ by the incoming radiation in the interval $[v_1, v_2]$ of the advanced
time, $v=t+|x|$, can be formally calculated by the limit
\begin{equation}\label{incoming}
\mathcal{E}_{\mathrm{in}}(v_{1},v_{2})=-\lim_{r\to +\infty}\int_{v_{1}}^{v_{2}}\int_{|x|=r}[S\cdot k](s-r,x)dx\,ds,
\end{equation}
where $k=x/|x|$ and $S$ is the Poynting vector, $S=(4\pi)^{-1}(E\wedge B)$.
It should be emphasized that (\ref{incoming}) is only a formal definition, since it is not known in general whether the above limit exists for a solution
of the relativistic Vlasov-Maxwell system.    
Analogously, the limit
\begin{equation}\label{outgoing}
\mathcal{E}_{\mathrm{out}}(u_{1},u_{2})=\lim_{r\to +\infty}\int_{u_{1}}^{u_{2}}\int_{|x|=r} [S\cdot k](s+r,x)dx\,ds
\end{equation}
gives the energy which is propagated to $\scri^+$ by the outgoing radiation in the interval $[u_1, u_2]$ of the retarded time, $u=t-|x|$.

A solution of the relativistic Vlasov-Maxwell system is isolated from incoming radiation if $\mathcal{E}_{\mathrm{in}}(v_{1},v_{2})=0$, for all 
$v_1, v_2\in\mathbb{R}$. In
\cite{C} it was proved that for small data this system admits solutions which satisfy this property. These solutions are defined by replacing the
Maxwell equations with the retarded part of the field. The resulting system has been called \textit{retarded} relativistic
Vlasov-Maxwell system and reads
\begin{equation}\label{vlasov}
\partial_{t}f_{\alpha}+\widehat{p}_{\alpha}\cdot\partial_{x}f_{\alpha}+q_{\alpha}(E_{\mathrm{ret}}+\widehat{p}_{\alpha}\wedge B_{\mathrm{ret}})
\cdot\partial_{p}f_{\alpha}=0,\quad\forall\,\alpha=1,...,N,
\end{equation}
\begin{equation}\label{field1}
E_{\mathrm{ret}}(t,x)=-\int_{\mathbb{R}^3}(\partial_{x}\rho+\partial_{t}j)(t-|x-y|,y)\,\frac{dy}{|x-y|},
\end{equation}
\begin{equation}\label{field2}
B_{\mathrm{ret}}(t,x)=\int_{\mathbb{R}^3}(\partial_{x}\wedge j)(t-|x-y|,y)\,\frac{dy}{|x-y|},
\end{equation}
where $\rho$ and $j$ are given by (\ref{source}). The purpose of this paper is to derive information about 
the asymptotic behaviour at future null infinity 
of such isolated  solutions, i.e. to study the properties of the outgoing radiation generated by the plasma.   
Let us first recall, for sake of reference, the global existence result of \cite{C}. Define 
\[
\mathcal{P}=\sup\{|p|:\,(x,p)\in\textrm{supp}f_\alpha(t),\, t\in\mathbb{R},\,1\leq\alpha\leq N\}
\]
and denote by $B_R(0)$ the sphere in $\mathbb{R}^6$ with center in the origin and radius $R>0$ and by $\lambda$ the set of constants
$\{q_\alpha,m_\alpha\}$. 

\begin{Theorem}\label{global}
Let $\fzero_{\alpha}(x,p)\geq 0$ be given in $C_{0}^{2}(\mathbb{R}^{3}\times\mathbb{R}^{3})$ such that 
$\fzero_{\alpha}=0$ for $(x,p,\alpha)\in B_{R}(0)^{c}\times\{1,...,N\}$. 
Define $\Delta=\sum_{\alpha}\sum_{|\mu|=0}^{2}\|\nabla^{\mu}\fzero_{\alpha}\|_{\infty}$,
where $\mu\in\mathbb{N}^{6}$ is a multi-index.
Then there exists a positive constant $\varepsilon=\varepsilon(R,\lambda)$ such that for $\Delta\leq\varepsilon$
the retarded relativistic Vlasov-Maxwell system has a unique global solution $\{f_{\alpha}\}\in (C^2)^N$ which satisfies 
$f_{\alpha}(0,x,p)=\fzero_{\alpha}(x,p)$. 
Moreover $F_{\mathrm{ret}}=(E_{\mathrm{ret}},B_{\mathrm{ret}})\in C^{1}(\mathbb{R}\times\mathbb{R}^{3})$ and there exists a positive 
constant $C=C(R,\lambda)$ 
such that $\mathcal{P}\leq C$ and the following estimates hold for all $\,(t,x)\in\mathbb{R}\times\mathbb{R}^3$:
\begin{equation}\label{estimate1}
|F_{\mathrm{ret}}(t,x)|\leq C\Delta (1+|t|+|x|)^{-1}(1+|t-|x||)^{-1},
\end{equation}
\begin{equation}\label{rhodecay}
\rho(t,x)\leq C\Delta (1+|t|+|x|)^{-3}.
\end{equation}
\end{Theorem}
The estimate (\ref{estimate1}) shows that the solution of theorem \ref{global} is isolated from incoming radiation in the sense specified above. 
We note that the statement of theorem \ref{global} differs from the main result of \cite{C} in two aspects. 
Firstly in \cite{C} only the case of
a single species of particles is considered. However the restriction to this case has been made only to simplify the notation and the
generalization of the result of \cite{C} to the case of a mixture is straightforward. Secondly we claim here that the distribution functions are 
twice continuously differentiable, whereas in \cite{C} we only proved that they are $C^1$. We shall return to this point at the end 
of the introduction.

We state now the main results of this paper. Let us define
\[
e(t,x)=\frac{1}{8\pi}\big (|E_{\mathrm{ret}}|^{2}+|B_{\mathrm{ret}}|^{2}\big )+\sum_{\alpha}\int_{\mathbb{R}^{3}}\sqrt{m_{\alpha}^{2}+
p^{2}}f_{\alpha}\,dp,
\]
\[
\mathfrak{p}(t,x)=\frac{1}{4\pi}(E_{\mathrm{ret}}\wedge B_{\mathrm{ret}})+\sum_{\alpha}\int_{\mathbb{R}^{3}} pf_{\alpha}\,dp,
\]
the local energy and momentum of a solution of (\ref{vlasov})--(\ref{field2}), respectively.
We also set 
\begin{equation}\label{bondi}
\mathcal{M}^{\vee}(u)=\int_{\mathbb{R}^{3}}[e-\mathfrak{p}\cdot k](u+|x|,x)\,dx
\end{equation} 
and note that $\mathcal{M}^{\vee}(u)$ is non-negative.

\begin{Theorem}\label{radiationfield}
Let $\{f_{\alpha}\}\in (C^2)^N$ be a solution of the retarded relativistic Vlasov-Maxwell system with data as stated in theorem 
\ref{global}, such that $F_{\mathrm{ret}}\in C^{1}(\mathbb{R}\times\mathbb{R}^{3})$ and
\begin{equation}\label{support}
f_\alpha(t,x,p)=0,\quad \forall x\in\mathbb{R}^3:\,|x|\geq R+a|t|,\,\,\alpha=1,...,N,
\end{equation}
for some $a\in [0,1)$.  
Then the limit
\begin{equation}\label{fieldlim}
F^{\mathrm{rad}}(u,k)=\lim_{|x|\to +\infty} |x|F_{\mathrm{ret}}(u+|x|,x) 
\end{equation}
exists and is attained uniformly in $k=x/|x|\in S^2$ and $u\in \mathcal{K}$, for all $\mathcal{K}\subset\mathbb{R}$ compact.
Moreover the radiation field $F^{\mathrm{rad}}=(E^{\mathrm{rad}},B^{\mathrm{rad}})$ satisfies the following algebraic properties: 
\begin{eqnarray*}
&E^{\mathrm{rad}}\cdot B^{\mathrm{rad}}=0,\quad |E^{\mathrm{rad}}|=|B^{\mathrm{rad}}|,\\
&E^{\mathrm{rad}}\cdot k=B^{\mathrm{rad}}\cdot k=0,\\
&(E^{\mathrm{rad}}\wedge B^{\mathrm{rad}})\cdot k=|E^{\mathrm{rad}}|^{2}.
\end{eqnarray*}
\end{Theorem}

\begin{Theorem}\label{bondimasslossformula}
Let $\{f_{\alpha}\}\in (C^2)^N$ be a solution of the retarded relativistic Vlasov-Maxwell system with data as stated in theorem \ref{global} and
such that $F_{\mathrm{ret}}\in C^{1}(\mathbb{R}\times\mathbb{R}^{3})$. Assume the following estimates hold
\begin{itemize}
\item[(i)] $\mathcal{P}\leq C$
\item[(ii)] $|F_{\mathrm{ret}}(t,x)|\leq C(1+|t|+|x|)^{-1}(1+|t-|x||)^{-1}$
\item[(iii)] $\rho(t,x)\leq C (1+|t|+|x|)^{-3}$
\end{itemize}
for some positive constant $C$ and for all $(t,x)\in\mathbb{R}\times\mathbb{R}^3$. Then $\mathcal{M}^{\vee}\in C^1$ and the 
following equation is satisfied:
\begin{equation}\label{lossformula}
\frac{d}{du}\, \mathcal{M}^{\vee}(u)=-\frac{1}{4\pi}\int_{S^{2}}|E^{\mathrm{rad}}(u,k)|^{2}dk.
\end{equation}
\end{Theorem}

Note that the conclusions of theorems \ref{radiationfield} and \ref{bondimasslossformula} apply to the solution of theorem
\ref{global}. However the proofs do not require the solution to be small. In fact the proofs of theorems 
\ref{radiationfield} and \ref{bondimasslossformula} do not make use explicitly of 
the Vlasov equation either. The only tools which enter
into play are the continuity equation, $\partial_t\rho + \nabla_x\cdot j=0$, and the local energy conservation law, $\partial_t
e+\nabla_x\cdot\mathfrak{p}=0$, which are of course satisfied by any ``good'' matter model. However since
the existence of global solutions of the retarded relativistic Vlasov-Maxwell system
which satisfy the assumptions of theorems \ref{radiationfield} and \ref{bondimasslossformula} is known by theorem \ref{global},
we restrict ourselves to consider this case.

Let us now comment the meaning of the results stated above.
In theorem \ref{radiationfield} it is claimed that the retarded field generated by the charge distribution is
asymptotically null and outwardly directed along the future pointing null geodesics. 
(We recall that an electromagnetic field $(E,B)$ is said to be \textit{null} if $E\cdot B=0$ and $|E|=|B|$, 
cf. \cite{S}, pag. 322.) This result follows essentially from \cite{GK}. 
However, for sake of completeness and to help the reader who is not familiar with the formalism used in \cite{GK}, 
we give in section 2 a complete proof of theorem \ref{radiationfield} adapted to our case. 
The property (\ref{support}), which for the solution of theorem \ref{global} follows from the estimate $\mathcal{P}\leq C$, is in turn a 
special case of the assumption made in \cite{GK} that the matter has to be contained in a timelike world-tube.

With regard to theorem \ref{bondimasslossformula}, it shows that the function $\mathcal{M}^{\vee}$ plays in the context of 
the retarded relativistic Vlasov-Maxwell system the same role as the Bondi mass in general
relativity (cf. \cite{B, W}). In fact, $\mathcal{M}^{\vee}$ is non-increasing and its variation on the interval $[u_1,u_2]$ equals the
energy  dissipated in form of outgoing radiation in such interval of the retarded time, as it follows from (\ref{outgoing}) and theorem
\ref{radiationfield}.
Although the Bondi mass loss formula in general relativity is extensively studied, it seems the first time that its generalization to 
plasma physics is
considered and that (\ref{lossformula}) appears in the literature. It should be mentioned, however, that a rigorous mathematical
derivation of the Bondi formula in general relativity is a much more difficult task (cf. \cite{ChKl, KlNi}). 

Let us now deal with the technical point concerning the regularity of the solution. We consider for simplicity the system for a 
single species of particle and denote by $f_{\mathrm{ret}}$ the unique global solution for small $C^2$ data. In \cite{C} we stated that 
$f_{\mathrm{ret}}\in C^1$; here we claim that $f_{\mathrm{ret}}\in C^2$.
This gain of regularity can be easily understood by appealing to the smoothing effect which
was pointed out in \cite{KS}. We recall that the solution of the Vlasov equation can be represented as
$f_{\mathrm{ret}}=f^{\mathrm{in}}(X(0),P(0))$ where $X(s), P(s)$ are the characteristics of (\ref{vlasov}) and are given by
\[
X(s)=x+\int_t^s \widehat{P}(\tau)\,d\tau,
\]
\[
P(s)=p+\int_t^s \Big (E_{\mathrm{ret}}(\tau,X)+\widehat{P}\wedge B_{\mathrm{ret}}(\tau,X)\Big )\,d\tau.
\]
It was observed in \cite{KS} that the time integral of the field evaluated on the characteristics is one derivative smoother than the
field itself provided that $X(s)$ is a timelike curve. The latter condition is satisfied by the solution of \cite{C} in virtue of the
estimate $\mathcal{P}\leq C$. Hence the characteristics are $C^2$ and since $f^{\mathrm{in}}$ is also given as a $C^2$ function, then the solution
of the Vlasov equation itself is twice continuously differentiable.

\section{Algebraic properties of the radiation field}
\setcounter{equation}{0}
In this section we prove theorem \ref{radiationfield}.
Let us denote by $\phi$ any of the components of the electromagnetic field, i.e. we set $\phi=E_i$ or $B_i$ and define
\[
F=\left\{ \begin{array}{ll}
-(\partial_t j_i+\partial_{x_i}\rho) & \textrm{ for the electric field }E,\\
(\partial_x\wedge j)_i & \textrm{ for the magnetic field }B.
\end{array} \right.
\]
Then $F\in C^{1}(\mathbb{R}\times\mathbb{R}^{3})$ and by means of (\ref{support}), $F(t,x)=0$ for  $|x|\geq R+a|t|$.
The retarded field defined by (\ref{field1}), (\ref{field2}) has the form
\[
\phi(t,x)=\int_{\Xi_a(t,x)}F(t-|x-y|,y)\,\frac{dy}{|x-y|},
\]
where $\Xi_a(t,x)=\{y\in\mathbb{R}^3:|y|\leq R+a|t-|x-y||\}$, which is a compact set for any fixed $t\in\mathbb{R}$, $x\in\mathbb{R}^3$. 
We have the following
\begin{Lemma}
\[
\lim_{|x|\to +\infty}|x|\,\phi(u+|x|,x)=\int_{\Omega_{a}(u)}F(u+k\cdot y,y)\,dy,
\]
uniformly in $(u,k)\in\mathcal{K}\times S^2$, for all $\mathcal{K}\subset\mathbb{R}$ compact, where
\[
\Omega_{a}(u)=\{y\in\mathbb{R}^{3}:|y|\leq (1-a)^{-1}(R+a|u|)\}.
\]
\end{Lemma}
\noindent\textit{Proof: }The crucial point to prove this lemma is that for $t=u+|x|$  and for $|x|$ large , i.e. where we need to evaluate the 
function $\phi$, the domain of integration $\Xi_a(t,x)$ is contained in $\Omega_{a}(u)$, a compact set whose measure depends only on the 
fixed $u$. Let $\rho=|x|^{-1}$ and define
\[
g(\rho)=[1-2\rho k\cdot y+\rho^{2}|y|^{2}]^{-1/2}=\frac{|x|}{|x-y|},
\]
\begin{eqnarray*}
h(\rho)
&=&
\rho^{-1}[1-\rho k\cdot y-(1-2\rho k\cdot y+\rho^{2}|y|^{2})^{1/2}]\\
&=&
|x|-k\cdot y-|x-y|,\quad\textrm{for } \rho\neq 0
\end{eqnarray*}
and put $h(0)=0$, $h'(0)=\frac{3}{2}(k\cdot y)^{2}-\frac{1}{2}|y|^{2}$, so that $h\in C^{1}(\mathbb{R})$ 
(here the prime denotes the derivative with respect to $\rho$). Then we have
\[
\lim_{|x|\to +\infty}|x|\,\phi(u+|x|,x)=\lim_{\rho\to 0}\int_{\Omega_{a}(u)}F(u+k\cdot y+h(\rho),y)\,g(\rho)\,dy. 
\]
Setting $G_{u,k}(\rho,y)=F(u+k\cdot y+h(\rho),y)g(\rho)$, we have to prove that
\[
\lim_{\rho\to 0}\Big|\int_{\Omega_{a}(u)} G_{u,k}(\rho,y)\,dy-\int_{\Omega_{a}(u)} G_{u,k}(0,y)\,dy\Big|=0,
\]
uniformly in $(u,k)\in\mathcal{K}\times S^2$.  
By the mean value theorem we have 
\[
G_{u,k}(\rho,y)=G_{u,k}(0,y)+\rho\, R_{u,k}(\rho,y),
\]
where the remainder is bounded as 
\[
|R_{u,k}(\rho,y)|\leq \sup\{|G'_{u,k}(\tau,y)|,\, 0\leq\tau\leq\rho\}.
\]
Hence
\begin{eqnarray*}
&&\Big| \int_{\Omega_{a}(u)} G_{u,k}(\rho,y)\,dy-\int_{\Omega_{a}(u)} G_{u,k}(0,y)\,dy\Big |\\
&&\leq\rho\,\sup\Big\{|G'_{u,k}(\tau,y)|,\, 0\leq\tau\leq\rho,\, y\in\Omega_a(u)\Big\}\textrm{Vol}[\Omega_a(u)].
\end{eqnarray*}
Since $G$ is $C^1$ and $\Omega_a(u)$ is compact, the lemma is proved. \prfe

By means of lemma 1, the radiation field is continuous and is given by 
\begin{equation}\label{radfield1}
E_{i}^{\mathrm{rad}}(u,k)=-\int_{\Omega_{a}(u)}(\partial_{i}\rho+\partial_{t}j_{i})(u+k\cdot y,y)\,dy,
\end{equation}
\begin{equation}\label{radfield2}
B_{i}^{\mathrm{rad}}(u,k)=\int_{\Omega_{a}(u)}(\partial_{x}\wedge j)_{i}(u+k\cdot y,y)\,dy.
\end{equation}
It remains to prove the algebraic properties of $(E^{\mathrm{rad}}, B^{\mathrm{rad}})$. In (\ref{radfield1}) we replace the identities 
\[
\partial_{t}j_{i}(u+k\cdot y,y)=\partial_{u}j_{i}(u+k\cdot y,y),
\]
\begin{eqnarray*}
\partial_{i}\rho(u+k\cdot y,y)
&=&\partial_{y_{i}}[\rho(u+k\cdot y,y)]+\partial_{y}\cdot [j(u+k\cdot y,y)]k_{i}\\
&&-\partial_{u}(j\cdot k)(u+k\cdot y,y)k_{i},
\end{eqnarray*}
the second one being a consequence of the continuity equation. After integrating by parts we get
\[
E_{i}^{\mathrm{rad}}(u,k)=\int_{\Omega_{a}(u)}\partial_{u}[(j\cdot k)k_{i}-j_{i}](u+k\cdot y,y)\,dy.
\]
Let $M$ denote the vector
\begin{equation}\label{Mdef}
M(u,k)=\int_{\Omega_{a}(u)}[(j\cdot k)k-j](u+k\cdot y,y)\,dy.
\end{equation}
Since the integrand function in (\ref{Mdef}) is $C^1$ and vanishes on the boundary of $\Omega_{a}(u)$, then we
have $E^{\mathrm{rad}}=\partial_uM$. Analogously, from (\ref{radfield2}) we get $B^{\mathrm{rad}}=\partial_{u}N$, where
\begin{equation}\label{Ndef}
N(u,k)=\int_{\Omega_{a}(u)}(j\wedge k)(u+k\cdot y,y)\,dy.
\end{equation}
The next lemma describes the algebraic properties of the vectors $M$, $N$.
\begin{Lemma}
$\forall (u,k)\in\mathbb{R}\times S^2$ the vector fields defined by (\ref{Mdef}), (\ref{Ndef}) satisfy
\begin{itemize}
\item[(1)] $M(u,k)\cdot k=N(u,k)\cdot k=0$
\item[(2)] $M(u,k)\cdot N(u,k)=0$
\item[(3)] $|M(u,k)|=|N(u,k)|$
\item[(4)] $(M(u,k)\wedge N(u,k))\cdot k=|M(u,k)|^{2}$.
\end{itemize}
\end{Lemma}
\noindent\textit{Proof: }The proof of (1) is straightforward. For (2) we put $j=j(u+k\cdot y,y)$ and $j'=j(u+k\cdot y',y')$ 
for short and write  
\begin{eqnarray*}
M\cdot N
&=&
\int \int [j-(j\cdot k)k]\cdot (j'\wedge k)\,dy'\,dy\\
&=&
\int \int j\cdot (j'\wedge k)\,dy'\,dy\\
&=&
-\int \int  j'\cdot (j\wedge k)\,dy'\,dy,
\end{eqnarray*}
where it is understood that the integrals are over the set $\Omega_a(u)$.
Then interchanging $y$ and $y'$ we get $M\cdot N=-M\cdot N$, i.e. $M\cdot N=0$.
To prove (3) we write
\begin{eqnarray*}
|M|^{2}
&=&
\int \int (j-(j\cdot k)k)\cdot(j'-(j'\cdot k)k)\,dy'\,dy\\
&=&
\int \int [j\cdot j'-(j\cdot k)(j'\cdot k)]\,dy'\,dy
\end{eqnarray*}
and
\[
|N|^{2}=\int \int (j\wedge k)(j'\wedge k)\,dy'\,dy.
\]
In the previous equation we use the following rule of vector calculus
\[
(a\wedge b)\cdot (c\wedge d)=(a\cdot c)(b\cdot d)-(a\cdot d)(b\cdot c),
\]
which is valid for any vectors $a$, $b$, $c$, $d$ and the identity (3) follows at once. To prove (4) we write
\begin{eqnarray*}
(M\wedge N)\cdot k
&=&
\int \int  [((j\cdot k)k-j)\wedge(j'\wedge k)]\cdot k\,dy'\,dy\\
&=&
-\int \int  [((j\cdot k)k-j)\wedge k]\cdot (j'\wedge k)\,dy'\,dy\\
&=&
\int \int  (j\wedge k)(j'\wedge k)\,dy'\,dy=|N|^{2}=|M|^{2}.
\end{eqnarray*}\prfe

The following lemma permits to relate the algebraic properties of the vectors $M$, $N$ to the ones of the radiation field and concludes the
proof of theorem \ref{radiationfield}. To simplify the notation we suppress the dependence on $k$ and denote by an upper dot the differentiation 
with respect to $u$.
\begin{Lemma}
Let $M(u), N(u)$ be $C^{1}$ vector fields satisfying the properties (1)--(4) of lemma 2. Then $\forall u\in\mathbb{R}$:
\begin{itemize}
\item[(a)] $\dot{N}(u)\cdot k=\dot{M}(u)\cdot k=0$
\item[(b)] $\dot{M}(u)\cdot\dot{N}(u)=0$
\item[(c)] $|\dot{M}(u)|= |\dot{N}(u)|$
\item[(d)] $(\dot{M}(u)\wedge\dot{N}(u))\cdot k=|\dot{M}(u)|^{2}$.
\end{itemize}
\end{Lemma}
\noindent\textit{Proof: }The identity (a) is proved at once by differentiating (1) with respect to $u$. To prove the other identities
we consider a coordinate system in which the $z-$axis is parallel to $k$. In this frame the vectors $M$ and $N$ have the form
\begin{eqnarray*}
&M(u)=(m_{1}(u),m_{2}(u),0),\\
&N(u)=(n_{1}(u),n_{2}(u),0).
\end{eqnarray*}
Now, because of (2), (3) and (4) of lemma 2:
\begin{eqnarray}
&m_{1}(u)n_{1}(u)+m_{2}(u)n_{2}(u)=0,\label{1}\\
&m_{1}(u)^{2}+m_{2}(u)^{2}=n_{1}(u)^{2}+n_{2}(u)^{2},\label{2}\\
&m_1(u)n_2(u)-m_2(u)n_1(u)=m_1(u)^2+m_2(u)^2.\label{3}
\end{eqnarray}
After some elementary algebra, (\ref{1})--(\ref{3}) give $m_{1}=n_{2}, \,m_{2}=-n_{1}$.
Hence the vectors $M,N$ can be represented in the following form:
\begin{eqnarray*}
&M(u)=(q(u),r(u),0)\Rightarrow \dot{M}=(\dot{q},\dot{r},0),\\
&N(u)=(-r(u),q(u),0)\Rightarrow \dot{N}=(-\dot{r},\dot{q},0),
\end{eqnarray*}
by which the properties (b), (c) and (d) follow at once. \prfe

\section{Bondi mass of the plasma}
\setcounter{equation}{0}
In this section we prove theorem \ref{bondimasslossformula}. Let us introduce
\begin{equation}\label{localbondi}
\mathfrak{m}^\vee(r,u)=\int_{|x|\leq r}[e-\mathfrak{p}\cdot k](u+|x|,x)\,dx.
\end{equation}
Note that $\mathfrak{m}^\vee(\cdot,u)$ is non-decreasing and so its limit as $r\to +\infty$ exists. We first prove (\ref{lossformula}) assuming that
\begin{itemize}
\item[($\sharp$)]
$\mathfrak{m}^\vee(r,u)$ converges as $r\to +\infty$ for all $u\in\mathbb{R}$. 
\end{itemize}
Assume ($\sharp$) holds. Then  $\mathcal{M}^{\vee}(u)$ is well defined as improper integral and we have 
\[
\mathcal{M}^{\vee}(u)=\lim_{r\to +\infty}\mathfrak{m}^\vee(r,u).
\]
Let $\mathcal{K}$ be a generic compact subset of $\mathbb{R}$. 
Evaluating the energy conservation law $\partial_{t}e=-\partial_{x}\cdot \mathfrak{p}$ on the future light cone 
corresponding to the value $u$ of the retarded time we have
\[
\partial_{u}e(u+|x|,x)=-\partial_{x}\cdot\mathfrak{p}(u+|x|,x). 
\]
In the previous equation we use the identity 
\[
\partial_{x}\cdot\mathfrak{p}(u+|x|,x)=\partial_{x}\cdot[(\mathfrak{p}(u+|x|,x)]-\partial_{u}(\mathfrak{p}\cdot k)(u+|x|,x)
\]
and so doing we get
\begin{equation}\label{cone}
\partial_{u}(e-\mathfrak{p}\cdot k)(u+|x|,x)=-\partial_{x}\cdot [\mathfrak{p}(u+|x|,x)].
\end{equation}
We now integrate (\ref{cone}) on the region $|x|\leq r$, and use the Gauss theorem to transform the right hand
side into a surface integral over the sphere of radius $r$. Since $f(t,x,p)$ is supported on the region $|x|\leq R+a|t|$, with $a\in [0,1)$  
then for $r$ large enough, $f(u+|x|,x,p)$ vanishes on $S_{r}=\{x:|x|=r\}$ and so we get
\begin{eqnarray}\label{coneint}
&&\lim_{r\to +\infty}\int_{|x|\leq r}\partial_{u}(e-\mathfrak{p}\cdot k)(u+|x|,x)\,dx\nonumber\\
&&\quad =-(4\pi)^{-1}\lim_{r\to +\infty}\int_{S_{r}}(E_{\mathrm{ret}}\wedge B_{\mathrm{ret}})\cdot k\,(u+r,x)\,dS_{r}.
\end{eqnarray}
The integral in the left hand side of (\ref{coneint}) is equal to
$\partial_u\mathfrak{m}^\vee(r,u)$, $\forall r>0$. 
By theorem \ref{radiationfield} and (\ref{coneint}) , $\partial_u\mathfrak{m}$ converges to the right hand side of
(\ref{lossformula}), uniformly in $u\in\mathcal{K}$, as $r\to +\infty$. Hence, using ($\sharp$) we infer that $\mathfrak{m}^\vee(r,u)$ 
converges uniformly in $u\in\mathcal{K}$ and also
\[
\lim_{r\to +\infty}\partial_u\mathfrak{m}^\vee(r,u)=\frac{d}{du}\Big(\lim_{r\to +\infty}\mathfrak{m}^\vee(r,u)\Big)=
\frac{d}{du}\mathcal{M}^{\vee}(u).
\]
Thus the proof of (\ref{lossformula}) is complete if we show that the property ($\sharp$) above is satisfied.
We note that to this purpose, a direct use of the
estimate (ii) in theorem \ref{bondimasslossformula} is not enough, since it entails $|E(u+|x|,x)|= O(|x|^{-1})$, as $|x|\to\infty$, which is too weak to bound 
$\mathcal{M}^{\vee}$. 
To solve this problem we rewrite $\mathcal{M}^{\vee}$ in a way that the decay at past null infinity, which is faster by means of
the absence of incoming radiation, enters into the estimates (in other words, we make the advanced time $v$ appear instead of the retarded
time $u$). For this purpose we introduce, besides $\mathfrak{m}^\vee(r,u)$, the following function
\begin{equation}\label{localbondi2}
\mathfrak{m}_\wedge(r,v)=\int_{|x|\leq r}(e+\mathfrak{p}\cdot k)(v-|x|,x)\,dx.
\end{equation}
We also set
\begin{equation}\label{bondi2}
\mathcal{M}_{\wedge}(v)=\lim_{r\to +\infty}\mathfrak{m}_\wedge(r,v).
\end{equation}
\begin{Lemma}\label{bondi2lemma}
Under the assumptions of theorem \ref{bondimasslossformula}, $\mathcal{M}_{\wedge}(v)$ is continuously differentiable and satisfies
\begin{equation}\label{bondi2cons}
\frac{d}{dv}\mathcal{M}_{\wedge}(v)= 0, \quad \forall v\in\mathbb{R}.
\end{equation}
\end{Lemma}
\noindent\textit{Proof: } 
We split $\mathcal{M}_{\wedge}(v)$ into four parts as follows:
\[
\mathcal{M}_{\wedge}(v)=I_{1}(v)+I_{2}(v)+I_{3}(v)+I_{4}(v),
\]
where
\[
I_{1}(v)=\frac{1}{8\pi}\int_{\mathbb{R}^3}  (|E_{\mathrm{ret}}|^{2}+|B_{\mathrm{ret}}|^{2})(v-|x|,x)\,dx,\]
\[I_{2}(v)=\sum_\alpha\int_{\mathbb{R}^3}\int_{|p|\leq C}\sqrt{m_\alpha^2+p^{2}} f_\alpha(v-|x|,x,p)\,dp\,dx,
\]
\[
I_{3}(v)=\frac{1}{4\pi}\int_{\mathbb{R}^3} (E_{\mathrm{ret}}\wedge B_{\mathrm{ret}})\cdot k\, (v-|x|,x)\,dx,\]
\[
I_{4}(v)=\sum_\alpha\int_{\mathbb{R}^3}\int_{|p|\leq C} p\cdot k\, f_\alpha(v-|x|,x,p)\,dp\,dx.
\]
Since $I_{3}$ (resp. $I_{4}$) is dominated by $I_{1}$ (resp. $I_{2}$), it suffices to estimate $I_{1}(v)$ and $I_{2}(v)$. 
By means of (ii), $I_1(v)$  is uniformly bounded in $v\in\mathcal{K}$. For $I_2(v)$ we use that $f_\alpha(v-|x|,x,p)=0$ for 
$|x|\geq R+a|v-|x||$,
which implies $f_\alpha(v-|x|,x,p)=0$ for $|x|\geq (1-a)^{-1}(R+a|v|)$. 
Thus 
\[
|I_{4}(v)|\leq C\textnormal{Vol}[\{x:|x|\leq (1-a)^{-1}(R+a|v|)\}]\leq C,\quad\forall v\in\mathcal{K}.
\] 
Since $\mathfrak{m}_\wedge(\cdot,v)$ is non-decreasing, the limit (\ref{bondi2}) exists for all $v\in\mathbb{R}$ and by the previous estimates,
$\mathcal{M}_{\wedge}(v)$ converges, as improper integral, uniformly in $v\in\mathcal{K}$.   
To prove (\ref{bondi2cons}) we repeat the argument which led to (\ref{cone}), now evaluating on the past light cone. So doing we get
\begin{equation}\label{cone2}
\partial_{v}(e+\mathfrak{p}\cdot k)(v-|x|,x)=-\partial_{x}\cdot [\mathfrak{p}(v-|x|,x)].
\end{equation}
Integrating and using the Gauss theorem we obtain,
\begin{eqnarray}\label{coneint2}
&&\lim_{r\to +\infty}\int_{|x|\leq r}\partial_{v}(e+\mathfrak{p}\cdot k)(v-|x|,x)\,dx\nonumber\\
&&\quad =-(4\pi)^{-1}\lim_{r\to +\infty}\int_{S_{r}}
(E_{\mathrm{ret}}\wedge B_{\mathrm{ret}})\cdot k\,(v-r,x)\,dS_{r},
\end{eqnarray}
where again we used that $f(v-r,x,p)$ vanishes on $S_r$ for large $r$.
The estimate (ii) implies that the right hand side of (\ref{coneint2}) tends to zero uniformly in $v\in\mathcal{K}$ and so we get
\begin{eqnarray*}
0&=&\lim_{r\to +\infty}\int_{|x|\leq r} \partial_{v}(e+\mathfrak{p}\cdot k)(v-|x|,x)\,dx\\
&=&\frac{d}{dv}\lim_{r\to +\infty}\int_{|x|\leq r} (e+\mathfrak{p}\cdot k)(v-|x|,x)=\frac{d}{dv}\mathcal{M}_{\wedge}(v),
\end{eqnarray*}
where the uniform convergence has been used to shift the derivative.
\prfe
Note that in the proof of lemma \ref{bondi2lemma} the estimate (iii) in theorem \ref{bondimasslossformula} has not been used. This will
become important for the completion of the proof of ($\sharp$).
The identity (\ref{bondi2cons}) represents the
counterpart of (\ref{lossformula}) on the backward cones of light. In fact, from one hand the conservation of $\mathcal{M}_{\wedge}(v)$ is 
due to the absence of incoming radiation and, on the other hand, the Bondi mass $\mathcal{M}^{\vee}(u)$ decreases as a consequence of the
emission of outgoing radiation.
 
We are able now to complete the proof of ($\sharp$). Integrating (\ref{cone2}) between $u$ and $u+2|x|$ we get
\[
(e+\mathfrak{p}\cdot k)(u+|x|,x)-(e+\mathfrak{p}\cdot k)(u-|x|,x)=-\int_{u}^{u+2|x|}\partial_{x}\cdot [\mathfrak{p}(v-|x|,x)]\,dv.
\]
Integrating in the region $|x|\leq r$ we get
\begin{eqnarray}\label{coneint3}
\int_{|x|\leq r}(e+\mathfrak{p}\cdot k)(u+|x|,x)\, dx
&=&\int_{|x|\leq r}(e+\mathfrak{p}\cdot k)(u-|x|,x)\, dx\\
&&-\int_{|x|\leq r}\int_{u}^{u+2|x|}\partial_{x}\cdot [\mathfrak{p}(v-|x|,x)]\,dv\,dx.\nonumber
\end{eqnarray}
Now we use the identity 
\begin{eqnarray*}
\partial_{x}\cdot [\int_{u}^{u+2|x|}\mathfrak{p}(v-|x|,x)\,dv]
&=&
2\mathfrak{p}\cdot k\,(u+|x|,x)\\
&&+\int_{u}^{u+2|x|}\partial_{x}\cdot[\mathfrak{p}(v-|x|,x)]\,dv.
\end{eqnarray*}
Substituting into (\ref{coneint3}) and using the Gauss theorem we obtain
\begin{equation}\label{bondiidentity}
\mathfrak{m}^\vee(r,u)=\mathfrak{m}_{\wedge}(r,u)+\mathfrak{q}(r,u).
\end{equation}
where
\begin{eqnarray*}
\mathfrak{q}(r,u)
&=&
-\int_{S_r}\int_{u}^{u+2r}\mathfrak{p}\cdot k\,(v-r,x)\,dv\,dS_r\\
&=&
-\frac{1}{4\pi}\int_{S_r}\int_u^{u+2r}(E_{\mathrm{ret}}\wedge B_{\mathrm{ret}})\cdot k\,(v-r,x)\,dv\,dS_r\\
&&-\sum_{\alpha}\int_{S_r}\int_u^{u+2r}\int_{|p|\leq C} p\cdot k\,
f_\alpha(v-r,x)\,dv\,dS_r.
\end{eqnarray*}

Using (ii) in the first term and (iii) in the second term, we conclude that $\lim_{r\to +\infty}|\mathfrak{q}(r,u)|$ is bounded 
for all $u\in\mathbb{R}$ and so the property ($\sharp$) follows from
(\ref{bondiidentity}) and lemma \ref{bondi2lemma}. This concludes the proof of theorem \ref{bondimasslossformula}.

\bigskip
\noindent
{\bf Acknowledgments:}
The results presented in this paper have been obtained while the author was preparing his PhD thesis at the Albert Einstein Institute in
Potsdam, which is thereby acknowledged for the hospitality. Support from the European Network HYKE (contract HPRN-CT-2002-00282) is also 
acknowledged.

\end{document}